\documentclass[preprintnumbers,amsmath,amssymb]{revtex4}
\usepackage{graphicx}
\usepackage{dcolumn}
\makeatletter
\input{epsf}
\begin{document}

\title{Noncommutative Geometry, Quantum Hall Effect and Berry Phase}

\author{B. Basu}
 \email{banasri@isical.ac.in}
  \author{P. Bandyopadhyay}
 \email{pratul@isical.ac.in}
\affiliation{Physics and Applied Mathematics Unit\\
 Indian Statistical Institute\\
 Kolkata-700108 }

\begin{abstract}
\begin{center}
{\bf Abstract}
\end{center}
 Taking resort to Haldane's spherical geometry we can visualize
fractional quantum Hall effect on the noncommutative manifold $M_4
\times Z_N$ with $N>2$ and odd. The discrete space leads to the
deformation of symplectic structure of the continuous manifold
such that the symplectic area is given by $\triangle p.\triangle
q=2\pi m \hbar$ with $m$ an odd integer which is related to the
Berry phase and the filling factor is given by $\frac{1}{m}$. We
here argue that this is equivalent to the noncommutative field
theory as prescribed by Susskind and Polychronakos which is
characterized by area preserving diffeomorphism. The filling
factor $\frac{1}{m}$ is determined from the change in chiral
anomaly and hence the Berry phase as envisaged by the star
product.
\end{abstract}
\maketitle

\section{Introduction}
It is well  known that the first quantized system of particles in
a strong magnetic field naturally realizes a noncommutative space.
We also know that a system of electrons moving in a strong
magnetic field exhibits certain interesting properties. When the
electron density lies at certain rational fractions of the density
corresponding to the fully filled Landau level, the electrons
condense into an incompressible fluid-like states whose
excitations are characterized by fractional charges and are
described by Laughlin wave functions. When we fill as many
electrons as possible in a two dimensional phase space, each
electron occupy an area $\triangle p. \triangle q~=~2 \pi \hbar$
which is responsible for the maximum electron density $n_e~=
~\frac{eB}{2\pi \hbar}$ in each Landau level. In the
noncommutative manifold $M_4 \times Z_N$ with $N>2$ and odd, the
symplectic area is deformed \cite{pb}. In this case, $\triangle p.
\triangle q~=~2 \pi m \hbar$ with m an odd integer and the
Laughlin states are achieved with $\nu=\frac{1}{m}$
\cite{sakita1,bb,pb3}. This deformation of symplectic structure
leads to the Berry phase as it corresponds to the association of
magnetic flux with an electron.

On the other side, Susskind \cite{sus} conjectured that the second
quantized noncommutative field theory can describe the Laughlin
states when we take into account the noncommutative version of the
$U(1)$ Chern-Simons theory. This also reveals the connection
between D brane physics with quantum Hall effect. The space
noncommutative condition requires infinite dimensional matrices
which was modified by Polychronakos \cite{poly} to realize
appropriate finite matrix model. However, in this framework
involving noncommutative field theory the system is characterized
by area preserving diffeomorphism.

In this note, we shall show that taking  resort to spherical
geometry the star product deformation of gauge fields effectively
leads to the change in chiral anomaly and hence induces the Berry
phase compatible with the fractional filling factor. This suggests
that the deformation of the symplectic structure of the
noncommutative manifold $M_4 \times Z_N$ with $N>2$ and odd
represents an equivalent description of the noncommutative version
of the $U(1)$ Chern-Simons theory which is characterized by area
preserving diffeomorphism. Thus we realize a dual relationship
between noncommutative manifold and noncommutative field theory.

\section{Noncommutative Manifold $M_4 \times Z_N$, Deformation of
Symplectic Structure and Berry phase}

Noncommutative geometry is characterized by the fact that
space-time structure acquires some fuzziness {\it i.e} points are
ill defined. The {\it fuzziness} of phase space variables may be
viewed such that the mean position of a particle at $q_\mu$ in the
external observable space has a stochastic extension given by a
stochastic variable $\hat{Q}_\mu$. The position and momentum of a
relativistic quantum particle in complexified space-time is then
given by
\begin{eqnarray}
  Q_\mu &~=&~q_\mu~+~i{\hat{Q}}_\mu \nonumber \\
 P_\mu &~=&~p_\mu~+~i\hat{P}_\mu
\end{eqnarray}
Introducing the dimensionless variable $\omega=\frac{\hbar}{lmc}$
it is shown that \cite{bpru} these relativistic canonical
commutation relations admit the following representations of
$\frac{Q_\mu}{\omega}$ and $\frac{P_\mu}{\omega}$ where the latter
are considered as acting on functions defined in phase space
\begin{eqnarray}\label{brook}
\frac{Q_\mu}{\omega} &=&-i(\frac{\partial}{\partial
p_\mu}+\phi_\mu) \nonumber  \\
\frac{P_\mu}{\omega} &= &-i(\frac{\partial}{\partial
q_\mu}+\psi_\mu)
\end{eqnarray}
where $\phi_\mu (\psi_\mu)$ are some matrix valued functions in
the noncommutative case.

In a recent paper \cite{pg} it has been pointed out that
noncommutative geometry having the space-time manifold $M_4 \times
Z_2 $ leads to the quantization of a fermion when the discrete
space-time is incorporated as an internal variable. Indeed, this
leads to the introduction of an anisotropic feature in the
internal space so that we can consider the space-time coordinate
in complex space-time as $z_\mu=x_\mu+i \xi_\mu$ where $\xi_\mu$
represents a {\it direction vector} attached to the space-time
point $x_\mu$. The two orientations of the {\it direction vector}
give rise to two internal helicities corresponding to fermion and
antifermion. The complex  space-time exhibiting the internal
helicity states can be written in terms of a two-component
spinorial variable $\theta (\overline{\theta})$ when the {\it
direction vector}  $\xi_\mu$  is associated with $\theta$ through
the relation $\xi_\mu=\frac{1}{2}{\lambda_\mu}^\alpha
\theta_\alpha$ $(\alpha= 1,2)$. This helps us to write the
relevant metric as $g_{\mu\nu}(x,\theta ,\overline{\theta})$. This
metric gives rise to the $SL(2,C)$ gauge theory \cite{ppa} where
the gauge fields ${\cal A}_\mu$ are matrix valued having the
$SL(2,C)$ group structure and the curvature ${\cal F}_{\mu\nu}$ is
given by
\begin{equation}
{\cal F}_{\mu\nu}=\partial_\mu {\cal A}_\nu -\partial_\nu {\cal
A}_\mu +[{\cal A}_\mu,{\cal A}_\nu]
\end{equation}
 The phase space variables may be written in terms of a gauge
 theoretical extension of a relativistic quantum particle. Indeed,
 from the relation (\ref{brook}) we can now identify
 $\phi_\mu(\psi_\mu)$ with gauge field ${\cal A}_\mu({\cal B}_\mu)$. That is, we
 can conceive of the position and momentum operators given by
\begin{eqnarray}\label{gauge}
\frac{Q_\mu}{\omega} &=&-i(\frac{\partial}{\partial
p_\mu}+{\cal A}_\mu) \nonumber \\
\frac{P_\mu}{\omega }&= &i(\frac{\partial}{\partial q_\mu}+{\cal
 B}_\mu)
\end{eqnarray}
It is now noted that the curvature ${\cal F}_{\mu\nu}$ will deform
the symplectic structure. Indeed, it will change the symplectic
form of the phase space as
\begin{equation}
\Omega=\frac{1}{2}g^{ij} dp_i\wedge dq_j
\end{equation}
with
\begin{equation}
g^{ij}=J^{ij}+\hbar\triangle^{ij}
\end{equation}
where
\begin{equation}
J^{ij}= \left[
\begin{array}{cc}
  0 & -I \\
  I & 0
\end{array}
 \right]
 \end{equation}
is associated with the usual symplectic structure and
$\triangle^{ij}$ is the curvature tensor coming from the induced
vector potential ${\cal A}_\mu$.

The Poisson bracket is modified as
\begin{eqnarray}\label{poison}
  \left\{f,g \right\}&=& g_{ij}\left( \frac{\partial f}{\partial q_i}\right)
  \left(\frac{\partial g}{\partial p_j}\right) \nonumber \\
       &=& \left\{f,g \right\} +\left\{f,g \right\}^{\prime}+...
\end{eqnarray}
where $f,g$ are arbitrary functions on the phase space, $g_{ij}$
is the inverse of $g^{ij}$. The Poisson bracket will help to
define the quantum field theory in the modified form if we replace
this by the commutator.

It is noted that in the manifold $M_4 \times Z_2$ when the
discrete space is considered as an internal extension, the
$direction$ $vector$ $\xi_\mu$ attached to the space-time point
$x_\mu \in M_4$, effectively corresponds to a vortex line. Again,
a vortex line is topologically equivalent to  a magnetic flux.
Thus in this picture, a fermion can be viewed as a boson attached
with a magnetic flux associated with the gauge field. In the
manifold $M_4 \times Z_N$ with $N>2$ and odd, we note that the
gauge field theoretical extension will induce a density of
vortices in the Bose field which corresponds to $\frac{1}{N}$ of a
vortex per boson. As an example, for $N$=3, we can view the system
such that each boson carries $\frac{1}{3}$ of magnetic flux
quantum. That is, three bosons will share one magnetic flux
quantum and the system will represent a three Bose particle
composite having one magnetic flux. It may be remarked here that
under duality relationship \cite{fish}, we can consider these
vortices as Bose particles and each hard core boson as a flux tube
carrying one Dirac flux quantum. This suggests that the ratio
between the number of new vortices and the number of new bosons
now becomes 3 to 1 and each new boson swallows three vortices.

When a $direction$ $vector$ $\xi_\mu$ is attached to a space-time
point $x_\mu$, the wave function of the particle concerned  will
be characterized by three Euler angles $\theta, \phi $ and $\chi$.
The angle $\chi$ corresponds to the rotational orientation around
the direction vector. The angular momentum relation is now  given
by
\begin{equation}\label{ang}
  \bf{J}=\bf{r} \times \bf{p} -\mu \hat{\bf{r}}
\end{equation}
Here $\mu$ corresponds to eigenvalue of the operator $i
\frac{\partial}{\partial \chi}$ having values $0,\pm
\frac{1}{2},\pm 1,\pm \frac{3}{2}$..... . In the ground state we
have $J=|\mu |$ and the wave functions become
\begin{equation}\label{wave}
  {{\psi}^J}_m~={D^J}_{m,\mu}(\theta,\phi,\chi)
\end{equation}
with $J=|\mu|$ and $m=-J,-J+1,...+J$. Indeed for
$J=m=|\mu|=\frac{1}{2}$, we can construct the following spinorial
functions
\begin{eqnarray}
u & =& cos\frac{\theta}{2}~ exp \left( \frac{i \phi+i
\chi}{2}\right)
  ={D^{\frac{1}{2}}}_{\frac{1}{2},-\frac{1}{2}}(\theta,\phi,\chi)
\nonumber   \\
v & =& sin\frac{\theta}{2} exp \left( \frac{-i \phi+i
\chi}{2}\right)
   ={D^{\frac{1}{2}}}_{\frac{1}{2},\frac{1}{2}}(\theta,\phi,\chi)
\end{eqnarray}
If we consider the geodesic projection coordinate
$$\zeta=\frac{v}{u}=tan\frac{\theta}{2}exp(-i \phi),$$
for the Hopf fibration $S^2=SU(2)/U(1)$, the base space turns to
be K\`{a}hler manifold and the symplectic structure is given by
\cite{hou}
\begin{eqnarray}\label{kah}
\Omega &=& 2i \frac{d\zeta \wedge
d\overline{\zeta}}{{(1+|\zeta|^2)}^2} \nonumber \\
 &=&
2i\frac{\partial ^2K}{\partial \zeta \partial
\overline{\zeta}}~d\zeta \wedge d\overline{\zeta}
\end{eqnarray}
where $K=ln (1+|\zeta|^2)$ is the K\`{a}hler potential. The
Hilbert space $\cal{H}_N$ on this Hopf fibration $S^2$ is composed
by the $N=2S+1$ one particle wave functions ${\psi^J}_m$ around
the Dirac monopole $\mu ~ (J=|\mu|)$. Evidently, for the three
particle system, we have $|\mu| ~=~\frac{3}{2}$ and the geodesic
projection coordinate is given by
$\zeta'=\left(\frac{v}{u}\right)^3$.  So from the relation
(\ref{kah}) we find that the symplectic structure $\Omega'$ in
this case is given by $\Omega'~=~3\Omega$. Thus the symplectic
structure for the noncommutative manifold $M_4 \times Z_3$
corresponds to the relation $\triangle p \triangle q~=~2 \pi m
\hbar$ with $m=3$. This can be generalized for the manifold $M_4
\times Z_N$ with $N>2$ and odd when we have $\triangle p \triangle
q~=~2 \pi m \hbar$ with $m=N$.

It is noted that in angular momentum relation (\ref{ang}), for
$\mu$ an integer we can have
\begin{equation}\label{intang}
\bf{J}=\bf{r} \times \bf{p} -\mu \hat{\bf{r}}=\bf{r}' \times
\bf{p}'
\end{equation}
 indicating the vanishing of magnetic field. This is the case for
the manifold $M_4 \times Z_N$ with $N>4$ and even. In this case,
the change in the symplectic structure will not be manifested
unless the state is split into a pair.

It is to be noted that the vector potential ${\cal A}_{\mu}$
responsible for the deformation of the symplectic structure
effectively gives rise to the geometrical phase of Berry. Indeed,
when a system $\hat{H}_0$ interacts with certain internal
Hamiltonian $\hat{h}_{int}$ to have
$\hat{H}_{eff}$=$\hat{H}_0$+$\hat{h}_{int}$, the response of the
internal system corresponding to the change of the external
canonical system becomes a part of the action as to associate the
change of field as \cite{kura}
\begin{equation}\label{kurat}
  S_{eff}=S_0~+ \hbar ~\Gamma(c)
\end{equation}
The physical importance of this geometrical phase $\Gamma(c)$ is
that it becomes a magnetic flux associated with the induced vector
potential ${\cal A}_\mu$. This implies that the factor $m$
occurring in the deformed symplectic structure $\triangle p
\triangle q~=~2 \pi m \hbar$ is associated with the Berry phase of
the system.

In the study of quantum Hall effect, we may visualize that the
external magnetic field realizes the noncommutative manifold $M_4
\times Z_N$ when we consider the electrons lie on the surface of a
3 dimensional sphere with a monopole at the center. In this
geometry, the deformation of the symplectic structure $\triangle p
\triangle q~=~2 \pi m \hbar$ leads to the fractional quantum Hall
effect when the filling factor is given by
$\nu=\frac{1}{m}=\frac{1}{2\mu}$  where $\mu$ is the monopole
strength \cite{bb,pb3}. It may be remarked here that the
deformation causes to increase the occupied area of an electron
and hence changes the electron density to lie at certain fractions
of the density corresponding to the density of the completely
filled Landau level. When these electrons condense into an
incompressible fluid we realize fractional quantum Hall effect.

\section{Noncommutative Field Theory, Chiral Anomaly and Berry
Phase}
 The noncommutative generalization for the free Maxwell Lagrangian
 density involves the star product of the noncommutative field
 strength $\hat{F}_{\mu\nu}$ constructed from the potential
 $\hat{A}_\mu$ and can be written as

\begin{equation}\label{field}
  \hat{F}_{\mu\nu}=\partial_\mu~\hat{A}_\nu~-\partial_\nu~\hat{A}_\mu
-ig~(\hat{A}_\mu* \hat{A}_\nu-\hat{A}_\nu~*~\hat{A}_\mu)
\end{equation}
and
\begin{equation}\label{lag}
  \hat{L}=-\frac{1}{4}\hat{F}_{\mu\nu}~*~\hat{F}^{\mu\nu}
\end{equation}
In terms of the conventional Maxwell tensor
$$F_{\mu\nu}
=\partial_\mu~A_\nu-\partial_\nu~A_\mu$$
it turns out that we can express $\hat{A_\mu}$ and
$\hat{F}_{\mu\nu}$ as follows \cite{jackiw}

\begin{equation}\label{jack}
  \hat{A}_\mu~=~A_\mu -\frac{1}{2}{\theta}^{\alpha\beta}A_\alpha
  (\partial_\beta~A_\mu~+~F_{\beta\mu})
  \end{equation}
\begin{equation}\label{jackiw}
\hat{F}_{\mu\nu}=F_{\mu\nu}+\theta^{\alpha\beta}F_{\alpha\mu}F_{\beta\nu}-
\theta^{\alpha\beta}A_\alpha \partial_\beta~F_{\mu\nu}
\end{equation}
where $g$ is absorbed in $\theta$. The tensor
$\theta^{\alpha\beta}$ is associated with the star product which
is defined as
\begin{equation}\label{star}
  (f*g)(x)=e^{\frac{1}{2}\theta^{\alpha\beta}}\partial_\alpha
  ~\partial_\beta~f(x)g(x')|_{x=x'}
\end{equation}
Apart from the total derivative term, the Lagrangian $\hat{L}$ is
given by
\begin{equation}\label{lagran}
  \hat{L}=-
  \frac{1}{4}F_{\mu\nu}F^{\mu\nu}+\frac{1}{8}\theta^{\alpha\beta}
  F_{\alpha\beta}
F_{\mu\nu}F^{\mu\nu}-\frac{1}{2}\theta^{\alpha\beta}F_{\mu\alpha}
F_{\nu\beta}F^{\mu\nu}+O(\theta^2)
\end{equation}
It is now well known that the star product effectively involves a
background magnetic field and so the second and third terms in
eqns. (\ref{jackiw}) and (\ref{lagran}) correspond the interaction
of this background field with the Maxwell field.

 We may recall
here that when a fermionic chiral current interacts with a gauge
field it gives rise to chiral anomaly which is given by
\begin{equation}\label{anomaly}
  \partial_\mu~{J_\mu}^5~=~\frac{1}{8 \pi^2}Tr~^{*}F_{\mu\nu}F^{\mu\nu}
\end{equation}
where ${J^5}_\mu $ is the axial vector current
$\overline{\psi}\gamma_\mu \gamma_5 \psi$ and
$$^*F_{\mu\nu}=\frac{1}{2}~\varepsilon^{\mu\nu\alpha\beta} ~F_{\alpha\beta}$$
 is the Hodge dual.
 The chiral anomaly is related to the Berry phase and we
have \cite{dbpb}
\begin{equation}\label{index}
  q=2\mu=-~\frac{1}{2}~\int(\partial_\mu {J_\mu}^5)~d^4x
  =-~\frac{1}{16 \pi^2}~\int Tr~^*F_{\mu\nu}F_{\mu\nu}~d^4x
\end{equation}
Here, $q$ is the Pontryagin index and $\mu$ corresponds to the
monopole strength.

 When we consider the
interaction of the chiral current with the noncommutative gauge
field having the gauge field strength
\begin{eqnarray}\label{change}
\hat{F}_{\mu\nu}&=& { F_{\mu\nu}
+\theta^{\alpha\beta}F_{\alpha\mu}F_{\beta\nu}-
\theta^{\alpha\beta}A_\alpha \partial_\beta~F_{\mu\nu}} \nonumber \\
 &=& F_{\mu\nu} + \widetilde{F}_{\mu\nu}
\end{eqnarray}
 we note that the chiral anomaly will be modified as
$$\frac{1}{8\pi^2}\left[ ~^*F_{\mu\nu}F_{\mu\nu}+~^*F_{\mu\nu}
\widetilde{F}_{\mu\nu}+~^*\widetilde{F}_{\mu\nu}\widetilde{F}_{\mu\nu}
\right]$$
 It is noted that we will have a change in the factor
$\mu$ which will induce a change in the Berry phase. The modified
value $\mu_{eff}$ will be given by
\begin{equation}\label{mono}
  \mu_{eff}=\mu+\mu^\prime+\tilde{\mu}=\mu+\overline{\mu}
\end{equation}
when we identify
\begin{equation}
  \mu=-\frac{1}{32\pi^2}~\int Tr~ ^*F_{\mu\nu}F_{\mu\nu}~d^4x,
\end{equation}
\begin{equation}
  \mu^\prime = - \frac{1}{32\pi^2}~\int Tr~^*F_{\mu\nu}\widetilde{F}_{\mu\nu}~d^4x,
\end{equation}
\begin{equation}
 \tilde{\mu}=-\frac{1}{32\pi^2}~\int Tr~^*\widetilde{F}_{\mu\nu}\widetilde{F}_{\mu\nu}~d^4x,
\end{equation}
This implies that the induced background magnetic field associated
with the star product effectively changes the number of magnetic
flux quanta in a specified area through its interaction with the
external gauge field strength which subsequently changes the
chiral anomaly and hence the Berry phase.

\section{Noncommutative Geometry and Quantum Hall Effect}
Susskind\cite{sus} has conjectured that quantum Hall effect can be
well described by the Chern-Simons action involving the
noncommutative version of the $U(1)$ gauge theory on a two
dimensional space. The action is given by
\begin{equation}
S=\frac{k}{4\pi}~\int d^3 x ~\epsilon^{\mu\nu\lambda}~({\hat
A}_\mu *
\partial_\nu {\hat A}_\lambda~+~\frac{2}{3}~{\hat A}_\mu *
{\hat A}_\nu * {\hat A}_\lambda)
\end{equation}
with $k=\frac{1}{\nu}$.

The theory is invariant under the noncommutative gauge
transformation
\begin{equation}
{\hat A}_\mu \rightarrow U^{-1}*{\hat
A}_\mu*U~+~i~U^{-1}*\partial_\mu U
\end{equation}
when $k$ is an integer.

The theory may be written in an equivalent form by choosing gauge
${\hat A}_0=0$ in which case the action becomes
\begin{equation}\label{action}
S=\frac{k}{4\pi}~\int d^3 x ~\epsilon^{ij} {\hat A}_i \partial_t
{\hat A}_j
\end{equation}
while the equation of motion for ${\hat A}_0$ must be imposed as a
constraint
\begin{equation}\label{field}
F_{ij}=\partial_i {\hat A}_j-\partial_j {\hat A}_i-i~({\hat
A}_i*{\hat A}_j-{\hat A}_j*{\hat A}_i)=0
\end{equation}
It turns out that this action and constraint also arise from a
matrix model given by
\begin{equation}
S=\frac{k}{\theta}~\int dt~ Tr\left(\frac{1}{2}~\epsilon_{ij}~D
X^i X^j \right)~+~k\int dt ~Tr(A)
\end{equation}
Where  $X$ and $A$ are Hermitian matrices. The action is invariant
under gauge transformation
\begin{eqnarray}
X^i & \rightarrow & U^{-1} X^i U  \\
A & \rightarrow & U^{-1} A U~+~iU^{-1} \partial_t U
\end{eqnarray}
as long as $U$ is taken to be trivial at $t=\pm \infty$ and $k$ is
an integer. In the gauge $A=0$, the action becomes
\begin{equation}
S=\frac{k}{\theta}~\int dt~ Tr\left(\frac{1}{2} ~\epsilon^{ij}
~X^i X^j\right)
\end{equation}
while the equation of motion for $A$ is
\begin{equation}
\left[X^1,X^2 \right]=i\theta
\end{equation}
which must be taken as a constraint. It is to be noted the
commutator here is just the matrix commutator. This has no
solution for finite dimensional matrices. A particular solution is
$X^i=y^i$ and $y^1$ and $y^2/\theta$ are the usual matrices
representing $x$ and $p$ in the harmonic oscillator basis.
Expanding the action and the constraint about the classical
solution $X^i=y^i+\theta \epsilon^{ij} {\hat A}_j$ where ${\hat
A}_j$ are functions of the noncommutative coordinates $y^i$, gives
precisely the Lagrangian(\ref{action}) with the
constraint(\ref{field}).

Polychronakos \cite{poly} has modified this formalism such that
the matrices $X^1$ and $X^2$ become finite dimensional. The
quantization of the inverse filling fraction and of the
quasiparticle number is shown to arise quantum mechanically and to
agree with the Laughlin theory.

In spherical geometry, where electrons lie on the surface of a
three dimensional sphere with a monopole at the centre, we can
consider the relation
\begin{equation}
\int_{M_4} F\wedge F~=~\int_{M_3} (A\wedge
dA~+~\frac{2}{3}~A\wedge A\wedge A)
\end{equation}

This suggest that in $3+1$ dimension we can replace the
noncommutative Chern-Simons term with the term
$^*\hat{F}_{\mu\nu}~*~\hat{F}_{\mu\nu}$. In fact, the
corresponding quantum Hall effect action in $3+1$ dimension is
given by
\begin{equation}\label{hall}
S~=~-\frac{\theta}{4}~\int ^*{\hat F}_{\mu\nu}~*~{\hat F}_{\mu\nu}
d^4x
\end{equation}
where $\theta$ is a coupling constant. So from the expression for
$\hat{F}_{\mu\nu}$ as given by (\ref{change}), we can write
\begin{equation}
^*\hat{F}_{\mu\nu}*\hat{F}_{\mu\nu}~=~^*{F}_{\mu\nu}~{F}_{\mu\nu}~
+~^*{F}_{\mu\nu}~\widetilde{F}_{\mu\nu}~+
~^*\widetilde{F}_{\mu\nu}~{\widetilde{F}}_{\mu\nu}~
\end{equation}
So the chiral anomaly is now changed and we have the associated
Berry phase factor as given by eqn.(\ref{mono}). Indeed the
expression for the Berry phase factor
\begin{equation}
\label{muef}
\mu_{eff}~=~\mu~+~\mu^\prime~+~\tilde{\mu}~=~\mu~+~\bar{\mu}
\end{equation}
implies that the term $\bar{\mu}$ is associated with the
modification caused by the background magnetic field inherent in
the star product formalism.

In an earlier paper \cite{bb1}, we have shown that in the lowest
landau level, we can have the filling factor $\nu$ in terms of the
Berry phase factor $\mu$ such that $\nu=\frac{1}{2\mu}$. In view
of this, we can now write
\begin{equation}\label{filling}
\nu_{eff}~=~\frac{1}{2\mu_{eff}}
\end{equation}
where $\mu_{eff}$ is given by (\ref{muef}).

This result can be compared with that of Dayi and Jellal
\cite{DaJ}. They have shown that if there is  noncommutativity of
coordinates in two dimensional space given by
\begin{equation}
\left[X,Y\right]~=~i\theta
\end{equation}
then there is a change in external magnetic field caused by the
inherent field in geometry and the effective field is given by
\begin{equation}
B_{eff}~=~\frac{B}{1-\frac{e\theta B}{4\hbar c}}
\end{equation}

This causes the change in the filling factor
\begin{equation}
\nu_{eff}~=~\frac{\phi_0 \rho}{B}~\left(1-\frac{e\theta B}{4\hbar
c}\right)
\end{equation}
where $\phi_0=hc/e$ and $\rho$ denotes electron density.

In sec.2 we have argued that in the noncommutative manifold $M_4
 \times Z_N$, the symplectic structure is modified such that
$\triangle p.\triangle q=2\pi m \hbar$ with $m$ an odd integer
given by $m=2\mu$. In relation to quantum Hall fluid, this implies
that we have now the filling factor given by
$\nu=\frac{1}{m}=\frac{1}{2\mu_{eff}}$. So from the relation
(\ref{filling}) we note that we arrive at the equivalent result in
terms of the noncommutative field theory. However, in the
noncommutative manifold $M_4 \times Z_N$, the Berry phase is
associated  with the deformation of symplectic structure where in
the framework of noncommutative field theory the system is
characterized by area preserving diffeomorphism. Thus we have a
duality relation between these two formalisms.

\section{Discussion}
It has been pointed out here that the noncommutative manifold $M_4
\times Z_N$ induces deformation of symplectic structure
 which leads to the fractional statistics for $N>2$ and odd. In
 fact, this deformation is related to the Berry phase factor which
 is responsible for the fractional filling factor of the
 incompressible Hall fluid. On the other hand, in spherical
 geometry where electrons reside on the surface of a sphere with a
 monopole at the center, the noncommutative field theory induces
 change in chiral
 anomaly leading to the Berry phase responsible for the fractional
 quantum Hall effect. However, the latter is characterized by
 area preserving diffeomorphism. Thus we can arrive at a dual
 relation between noncommutative manifold $M_4 \times Z_N$ (with
 $N>2$ and odd) and noncommutative field theory.

It may be recalled that there exists a correspondence between
noncommutative geometry and quantum space which is characterized
by certain quantum group symmetry and both represent a lattice
structure. Kogan \cite{kogun} has demonstrated the connection
between quantum symmetry, magnetic translation and area preserving
diffeomorphism in Landau problem and discussed the relevance of
$U_q(SL(2))$ in quantum Hall system. In a recent paper \cite{pb3}
a possible link between the Berry phase factor $\mu$ with the
deformation parameter $q$ of the deformed algebra $U_q(SL(2))$ has
been suggested. Indeed, Kogan's analysis suggests that in a
quantum Hall system the deformation parameter $q$ of the symmetry
$U_q(SL(2))$ is related to the filling factor $\nu$ of the form
$1/m$, $m$ being an odd integer, through the relation $q=exp(i2\pi
\nu)$ which implies that $q$ is related to the Berry phase factor
$\mu$ through the relation $q=exp(i\frac{2\pi}{2 \mu_{eff}})$. It
may be observed that the association of fractional quantum Hall
states with the $Z_p$ spin system \cite{spin} directly links up
the quantum space and noncommutative geometry with the lattice
structure. In fact for $q=exp(\frac{2\pi i}{3})$,
$q=exp(\frac{2\pi i}{5})$,...... we have the corresponding
$Z_3,Z_5$.... spin system representing fermion number
$\frac{1}{3}, \frac{1}{5}$ which corresponds to the Berry phase
factor $\mu_{eff}=\frac{3}{2},\frac{5}{2}$ and so on. This
evidently relates the deformation parameter of the quantum group
$U_q(SL(2))$ when $q$ is a root of unity with noncommutative
geometry which realizes these Berry phase factors either through
the deformation of the symplectic structure in the noncommutative
manifold $M_4 \times Z_N$ or through the change in chiral anomaly
in the noncommutative field theory.

\end{document}